\documentclass[trackchanges,twocolumn]{aastex62}
\usepackage{amsmath,amstext}
\usepackage{units}
\usepackage{url}

\newcommand{\E}[1]{\times10^{#1}}
\newcommand{\msol}{ \, M_\sun }

\newcommand{\bi}{\begin{itemize}}
\newcommand{\ei}{\end{itemize}}
\newcommand{\commentOut}[1]{}

\shortauthors{Shen, Quataert, \& Pakmor}

\begin{document}

\title{\bf \Large{The Progenitors of Calcium-Strong Transients}}

\author[0000-0002-9632-6106]{Ken J. Shen}
\affiliation{Department of Astronomy and Theoretical Astrophysics Center, University of California, Berkeley, CA 94720, USA}

\author[0000-0001-9185-5044]{Eliot Quataert}
\affiliation{Department of Astronomy and Theoretical Astrophysics Center, University of California, Berkeley, CA 94720, USA}

\author[0000-0003-3308-2420]{R\"{u}diger Pakmor}
\affiliation{Max-Planck-Institut fur Astrophysik, Karl-Schwarzschild-Str.\ 1, D-85741 Garching, Germany}

\correspondingauthor{Ken J. Shen}
\email{kenshen@astro.berkeley.edu}

\begin{abstract}

A new class of faint, spectroscopically peculiar transients has emerged in the last decade.  We term these events ``calcium-strong transients'' (CaSTs) because of their atypically high calcium-to-oxygen nebular line ratios.  Previous studies have struggled to deduce the identity of their progenitors due to a combination of their extremely extended radial distributions with respect to their host galaxies and their relatively high rate of occurrence.  In this work, we find that the CaST radial distribution is consistent with the radial distribution of two populations of stars: old (ages $ > 5 \, {\rm Gyr}$), low-metallicity ($Z/Z_\odot < 0.3$) stars and globular clusters.  While no obvious progenitor scenario arises from considering old, metal-poor stars, the alternative production site of globular clusters leads us to narrow down the list of possible candidates to three binary scenarios: mergers of helium and oxygen/neon white dwarfs; tidal disruptions of helium white dwarfs by neutron stars; and stable accretion from low-mass helium-burning stars onto white dwarfs.  While rare in the field, these binary systems can be formed dynamically at much higher rates in globular clusters.  Subsequent binary hardening both increases their interaction rate and ejects them from their parent globular clusters prior to mass transfer contact.  Their production in, and ejection from, globular clusters may explain their radial distribution and the absence of globular clusters at their explosion site.  This model predicts a currently undiscovered high rate of CaSTs in nuclear star clusters.  Alternatively, an undetermined progenitor scenario involving old, low-metallicity stars may instead hold the key to understanding CaSTs.

\end{abstract}



\section{Introduction}
\label{sec:intro}

The past two decades have seen a rapid increase in the number of observed classes of explosive transients beyond the standard Type Ia and core-collapse supernova (SN) types (for a recent review, see \citealt{taub17a}).  In this work, we focus our attention on the ``Ca-rich SNe'' \citep{pere10} or ``Ca-rich gap transients'' \citep{kasl12}, which have earned their name due to the strength of their Ca nebular features as compared to their O features and their typical peak magnitudes in the gap between those of SNe and classical novae.  However, while these transients do exhibit uniquely high nebular line ratios of Ca to O, this does not necessarily require that Ca is produced in large quantities because it is such an effective coolant.  An equally plausible explanation is that these objects are O-poor.  Estimates of the relative abundances of various isotopes produced by these explosions require suites of detailed reactive hydrodynamic and radiation transport simulations, which have yet to be fully undertaken (for preliminary explorations, see \citealt{pere10} and \citealt{dess15a}).  Thus, in this paper, we refer to the members of this class as ``Ca-strong transients'' (CaSTs).

SN 2005E, the prototype CaST, was discovered by the Lick Observatory Supernova Search and first analyzed by \cite{pere10}.  It peaked at $M_R \simeq -15.5$ and evolved relatively rapidly, decreasing by $\sim \unit[0.7]{mag}$ in $B$-band 5 days from peak.  Its photospheric spectra showed He features and no sign of H; however, the most interesting spectral features were only revealed months after explosion, when its nebular spectrum showed an anomalously high ratio of Ca to O, $5-10$ times higher than typical Type Ib SNe.  SN 2005E's location with respect to its host galaxy also proved to be extremely interesting: it occurred $\unit[23]{kpc}$ from the galaxy nucleus, well outside the locations of the vast majority of other SNe and even short gamma-ray bursts.

At present, there are seven other candidate members of the CaST class that share similar photometric and spectroscopic properties to SN 2005E, as well as six others that lack good photometric coverage and four more that share some similarities to the class but also possess important differences.  These objects are summarized in Table \ref{tab:summary}.  Throughout this paper, we will refer to the subset of the 8 best candidates as the gold sample and the larger set of 14 candidates as the silver sample.  We exclude the four least similar objects (PTF09dav, iPTF15eqv, SN 2016hnk, and SN 2019bkc) from our analysis in this work.

\begin{deluxetable*}{cccccc}
\label{tab:summary}
\tablecaption{Summary of known CaSTs and candidates}
\tablehead{
\colhead{SN name} & \colhead{Peak $R$ mag.}  & \colhead{Comments} & \colhead{Gold} & \colhead{Silver} & \colhead{References}  }
\startdata
SN 2005E		& $-15.5$	& & \textbullet &\textbullet & 1 \\
SN 2007ke	& $-16.3$	& Longer rise to peak than others & \textbullet &\textbullet & 1, 2, 3 \\
PTF10iuv / SN 2010et & $-16.0$ & & \textbullet&\textbullet & 3 \\
PTF11bij		& $-15.9$	&  &\textbullet &\textbullet & 3 \\
PTF11kmb	& $-15.5$	& & \textbullet& \textbullet& 4 \\
PTF12bho		& $-16.1$	& Unique photospheric spectrum & \textbullet& \textbullet& 4 \\
SN 2012hn	& $-15.7$	&  & \textbullet&\textbullet & 5 \\
iPTF16hgs	& $-15.6$\tablenotemark{a} &  Double-peaked light curve & \textbullet&\textbullet & 6 \\
\hline
SN 2000ds	&$<-13.9$\tablenotemark{b} &  Peak unconstrained; no near-max.\ spectrum & & \textbullet& 1, 3, 7 \\
SN 2001co	& $<-15.9$\tablenotemark{b} &  Peak unconstrained & & \textbullet& 1, 3, 8 \\
SN 2003H		& $<-15.3$\tablenotemark{b} & Peak unconstrained & & \textbullet& 1, 3, 8 \\
SN 2003dg	& $<-15.5$\tablenotemark{b} & Peak unconstrained; no near-max.\ spectrum & & \textbullet& 1, 3, 8 \\
SN 2003dr	& $<-14.9$\tablenotemark{b} &  Peak unconstrained; no near-max.\ spectrum & & \textbullet& 1, 3, 8 \\
SN 2005cz	& $<-15.3$ & Peak unconstrained & & \textbullet& 9 \\
\hline
PTF09dav		&  $-16.3$	 & Sc, Sr, nebular H; lower velocities & & & 10 \\
iPTF15eqv	& $<-15.4$\tablenotemark{b} &  Strong H; peak unconstrained & & & 11 \\
SN 2016hnk	& $<-16.2$\tablenotemark{c} & No H, but similar to PTF09dav & & & 12, 13 \\
SN 2019bkc	& $-17.3$ & Brighter peak mag.; rapidly declining light curve & & & 14, 15
\enddata
\tablenotetext{a}{Magnitude of the second peak.}
\tablenotetext{b}{Unfiltered magnitude.}
\tablenotetext{c}{ATLAS cyan filter.}
\tablecomments{References:
1.\ \cite{pere10}; 
2.\ \cite{chu07a};
3.\ \cite{kasl12};
4.\ \cite{lunn17a};
5.\ \cite{vale14a};
6.\ \cite{de18a};
7.\ \cite{puck00a};
8.\ \cite{fili03a};
9.\ \cite{kawa10a};
10.\ \cite{sull11};
11.\ \cite{mili17a};
12.\ \cite{tonr16a};
13.\ \cite{sell18a};
14.\ \cite{chen19a}.
15.\ \cite{pren19a}.}
\end{deluxetable*}

While the number of CaSTs is fairly small at present, there are still ubiquitous properties that must be matched by any viable explanation.  Most CaSTs show He features in their photospheric spectra, along with varying degrees of O, Mg, Si, Ca, and Fe \citep{kasl12}.  The presence of these elements, the lack of H, and the photospheric velocities of $\sim \unit[10^4]{km \, s^{-1}}$ suggest explosive burning of an evolved star or white dwarf (WD).  At nebular phases, as previously discussed, they exhibit atypically high Ca/O ratios, so the ejecta likely do not contain large amounts of O.

CaSTs reach absolute peak magnitudes of $M_R=-15$ to $-16.5$ and remain within 50\% of this peak for $\unit[10-15]{d}$.  The light curve maxima and shapes imply ejecta masses of $ 0.1-1.0 \msol$ and radioactive power sources composed of $\sim 0.01 \msol$ of $^{56}$Ni, $^{48}$Cr, or some combination of the two.  These properties are also suggestive of explosive burning of an evolved star or WD.

The CaST rate has been estimated to be $ \unit[1.21^{+1.13}_{-0.39} \times 10^{-5}]{events \, yr^{-1} \, Mpc^{-3}}$ \citep{froh18a}, which is quite high ($33-94$\% of the Type Ia SN rate) given that the first paper on these transients was only published a decade ago.  While the rate is still quite uncertain, it is clear that the progenitor system and explosion mechanism cannot be very rare.

Finally, and perhaps most constraining, SN 2005E's large host galaxy radial offset has proven to be an enigmatic property of the whole class, with an average offset of $39$ and $\unit[24]{kpc}$ for the gold and silver samples, respectively.  Furthermore, no obvious signs of star formation or globular clusters (GCs) exist at the sites of the CaSTs that have been observed deeply enough to place meaningful constraints (\citealt{lyma14a,lyma16a,lunn17a}; although see \citealt{lunn17a} for a possible host detection for PTF11kmb).  The host galaxies of the observed CaSTs show a possible preference for elliptical galaxies in clustered environments \citep{kasl12,fole15b,lunn17a}: 6 of the 8 gold-sample hosts have more than 10 nearby galaxy neighbors, and 7 of the 8 gold-sample hosts are E/S0 galaxies, although this fraction decreases to 9 of the 14 silver-sample hosts.

In addition to representing a new and as yet unexplained astrophysical phenomenon, CaSTs may also play an important role in chemical enrichment.  If the Ca yield per transient is indeed as large as $\sim 0.1 \msol$, as suggested by \cite{pere10}, their high rate implies a significant, perhaps even dominant, contribution to the Ca abundance in clusters \citep{mulc14a,mern16a}; without this contribution, the observed cluster Ca abundance is higher than can be explained by standard production in core-collapse and Type Ia SNe \citep{depl07a}.

In this work, we summarize the existing models in the literature and evaluate them with respect to the observed constraints.  Our favored models invoke binaries containing at least one WD: He + O/Ne WD binaries, He WD + neutron star (NS) binaries, and low-mass He-burning star + WD binaries.  We show that the galactocentric radial distribution of globular clusters (GCs) is consistent with that of the CaSTs.  Since no GC has yet been discovered at the site of a CaST, we postulate that CaST progenitors are formed dynamically in GCs at rates competitive with that in the field and then ejected from their parent GCs via binary hardening recoil kicks prior to interacting.   Because escape speeds from GCs are lower than their galactocentric velocities, the ejected CaST progenitors will inherit the GC radial distribution.  However, while dynamical formation in, and ejection from, GCs is consistent with the CaST radial distribution, this explanation suffers from the relatively small amount of mass available in GCs.  We also find that the CaST radial distribution is consistent with that of old (ages $> 5 \, {\rm Gyr}$), low-metallicity ($Z/Z_\odot <0.3$) stars in the local universe; however, a progenitor scenario restricted to such stars is not obvious.


\section{Proposed progenitor models}

In this section, we describe the various models that have been proposed in the literature and compare and contrast their characteristics with observed properties of the CaST class.


\subsection{Ultra-stripped-envelope core-collapse supernovae}

Stripped-envelope core-collapse SNe (Type Ib/c SNe) lack H features due to the removal of most or all of their H envelopes prior to explosion and evolve more rapidly than typical core-collapse SNe \citep{fili97a}.  However, with $^{56}$Ni masses $\sim 0.1 \msol$ and ejecta masses $\sim 2 \msol$, Type Ib/c SNe are still brighter and evolve on longer timescales than CaSTs \citep{lyma16b}.  Recent work has begun to explore even more extreme mass loss, which may lead to ultra-stripped SNe \citep{taur13a,mori17a,de18b} with $^{56}$Ni masses and ejecta masses of only $\sim 0.03$ and $\sim 0.1 \msol$, respectively, yielding rapidly evolving light curves that can match those of CaSTs.

\cite{kawa10a} suggested such a low-mass stripped-envelope core-collapse SN origin for SN 2005cz, a member of our silver sample but not our gold due to the lack of photometry near peak.  Although SN 2005cz's host is an elliptical galaxy, it shows evidence for relatively recent star formation $\unit[10^7-10^8]{yr}$ ago, which allows for the possibility that SN 2005cz arose from a young star.

However, while SN 2005cz's host galaxy as a whole may have a young stellar population, \cite{pere11} found no evidence for star formation at the specific site of SN 2005cz.  Moreover, such a scenario does not explain the CaST population in general due to their large average offsets from their host galaxies in addition to their galaxies' lack of obvious star formation \citep{lyma13a,lyma14a,lyma16a}.  We thus do not consider ultra-stripped core-collapse SNe to be a viable explanation for the CaSTs.


\subsection{Tidal disruptions of white dwarfs by intermediate-mass black holes}

WDs that pass close enough to black holes with masses $\lesssim 10^5 \msol$ (intermediate-mass black holes; IMBHs) will become tidally disrupted and compressed as they approach pericenter \citep{ross08a,ross09b}.  For deeply plunging WDs, the compression is strong enough to ignite thermonuclear burning, which will give rise to an optical transient \citep{macl16a}.  \cite{sell15a} proposed that this mechanism might explain the CaST class, invoking interactions with IMBHs formed inside of GCs or dwarf galaxies.

Although the initial studies of \cite{ross08a,ross09b} used a 7-isotope nuclear network that did not include Ca, recent studies have explored the nucleosynthesis in more detail.  \cite{tani18a} followed the disruption and detonation of a $0.45 \msol$ He WD and found that $0.3 \msol$ of $^{56}$Ni was produced, far in excess of what is implied from the faint peak luminosities of CaSTs.  This trend was also confirmed by \cite{kawa18a} for a larger set of WD masses, penetration factors, and IMBH masses.  \cite{anni18a} did find that Ca-dominated debris can be created for less deeply plunging disruptions.  However, in general, it appears difficult to avoid synthesizing relatively large amounts of radioactive isotopes in WD tidal disruptions by IMBHs, which implies light curves that do not match those of CaSTs.  

Furthermore, there is still no direct evidence for the existence of IMBHs (see \citealt{sell15a} for a discussion).  Moreover, \cite{sell15a} searched for and did not detect X-ray emission in a candidate CaST (SN 2012hn) that would be expected from subsequent accretion of bound material onto the IMBH.  We note that \cite{sell18a} also found no X-ray emission at the site of SN 2016hnk, but we do not include this transient in our gold or silver sample due to its similarity to PTF09dav.

The strongest evidence against this scenario comes from the non-detections of GCs or dwarf galaxies at the sites of several CaSTs.  Unlike the stars in the binary scenarios described next, IMBHs cannot be ejected from their GCs or dwarf galaxies, and so their associated CaSTs should occur on top of GC or dwarf galaxy hosts.  For these reasons, we consider this scenario an unlikely explanation for the CaSTs.


\subsection{Helium shell explosions on white dwarfs}

The first explanation proposed for the prototype SN 2005E was a detonation in a He shell accreted onto a WD from a He WD or a He-burning star \citep{pere10}.  However, the detonation of a $0.3 \msol$ He envelope, as proposed by \cite{pere10}, produces a large amount of $^{56}$Ni \citep{shen10,wk11} unless it is strongly enriched by C or occurs at low density \citep{wald11,sim12}.  One possible solution occurs in a merging double WD system, in which much of the unstably transferred He does not reach high densities because it is virialized by shocks when it directly impacts the accreting WD's surface; in this case, the overproduction of $^{56}$Ni might be avoided \citep{dess15a}.

It is also necessary that the He shell detonation does not trigger a core detonation, as this would lead to a double detonation Type Ia SN \citep{nomo82b,fhr07,fink10,poli19a,town19a}.  Such an outcome can be avoided if the accreting WD has a mass $ \lesssim 0.8 \msol$ or is composed of O/Ne \citep{shen14a}, as expected for WDs with initial masses $\gtrsim 1.15 \msol$.  Thus, the merger of a He + O/Ne WD may produce a CaST: detonation of the relatively low-density disrupted He WD may avoid overproduction of radioactive isotopes, and the O/Ne WD core precludes a double detonation Type Ia SN.

Another alternative is that the burning in the He envelope takes place as a subsonic deflagration that never transitions into a supersonic detonation.  Without such a deflagration-to-detonation transition,\footnote{Deflagration-to-detonation transitions were long assumed to occur in near-Chandrasekhar-mass convective C-burning WDs, leading to Type Ia SNe \citep{khok91a}.  However, the connection of the recently discovered peculiar Type Iax SNe to models of pure deflagration explosions with no detonation transition \citep{bran04a,fink14a,long14a} suggests that deflagration-to-detonation transitions may not occur under astrophysical conditions.} He shell deflagrations are natural outcomes of inefficient He shell convection, which occurs in the $ \sim 0.2 \msol$ envelopes that are transferred from $\sim 0.5 \msol$ non-degenerate He-burning stars onto WD accretors \citep{ergm90a,yung08}.\footnote{In principle, inefficient He shell convection may also be achieved in double WD binaries with He WD donors if they undergo stable mass transfer \citep{bild07,sb09b,shen10}.  However, \cite{shen15a} found that essentially all double WD binaries undergo dynamically unstable mass transfer, which explains the mismatch between the theoretical formation rate of such double WD binaries and the observed AM Canum Venaticorum population \citep{nele01b,cart13a,brow16b}.}  The slower flame speed of a deflagration allows for pre-expansion of the fuel ahead of the burning front, so that the burning proceeds at lower densities and less $^{56}$Ni is produced.

To date, there have only been two WD He shell deflagration simulations performed \citep{wk11}, and only in spherical symmetry, which is a rough approximation given the inherent multi-dimensionality of a shell deflagration.  With these caveats, He shell deflagrations do seem to produce an interesting mix of isotopes possibly consistent with CaSTs, with ejecta mass fractions of $0.02-0.06$ for radioactive $^{48}$Cr and $\simeq 10\%$ for $^{40}$Ca (the O yields were not reported by \citealt{wk11}), thus avoiding the nucleosynthetic issues of detonations in high-density He shells.

Much work remains to be done, but at present, He shell detonations or deflagrations remain viable CaST explosion mechanisms.  However, they must also satisfy the large galactocentric radial distribution constraint of the observed CaST population.  As we discuss in Section \ref{sec:gc}, GCs can match this radial distribution, so if the formation and / or interaction rates of the relevant He-accreting WD binaries are negligible in the field but are significantly enhanced inside of GCs, the radial distribution could match that of the CaSTs.  This requirement appears to rule out He + C/O WD mergers, which have an interaction rate in the field of a Milky Way-like galaxy of $\unit[0.003]{yr^{-1}}$ \citep{nele01b,meng15a,brow16b}, $\sim 10\%$ of the Type Ia SN rate; these may instead yield R Corona Borealis stars \citep{clay96} and Type Ia SNe via double detonations \citep{guil10,pakm13a,shen18a,shen18b}.  However, there are indications that He + O/Ne WD binaries are much rarer in the field than other combinations of double WDs (Toonen 2019, private communication).  If they are instead formed frequently in GCs by dynamical processes, they could be possible CaST progenitors.  Similarly, the low-mass non-degenerate He-burning star + WD binaries that may yield He shell deflagrations are relatively rare in the field \citep{geie13a} but may also be overproduced in GCs.

As previously noted, there is no evidence for GCs at the sites of the CaSTs with deep enough limits \citep{lyma14a,lyma16a,lunn17a}.  At face value, this seems to rule out GCs as formation sites of CaSTs.  However, dynamical processes inside of GCs will yield recoil kicks to these binaries that may eject them from the GCs, after which gravitational wave radiation decreases their separations and causes them to interact; we explore this possibility in Section \ref{sec:mechanism}.  Future work focused on the dynamical formation and evolution of these specific binaries in GCs is required, but at present, He shell detonations during He + O/Ne WD mergers or deflagrations from low-mass He-burning star + WD binaries remain plausible CaST progenitor scenarios.


\subsection{Tidal disruptions of white dwarfs by neutron stars}

The tidal disruption of a WD by a NS was first examined in detail by \cite{pasc09a,pasc11a}, although they did not include the important effects of nuclear burning.  Recent work has expanded on these initial explorations by including nuclear reactions and focusing on the outflows driven from the accretion disk formed by the disrupted WD \citep{metz12a,fern13a,marg16a,fern19a,zena19a}.  These studies find that as material viscously spreads from the disruption radius $\sim \unit[10^9]{cm}$ inward to the NS, the midplane temperature becomes hotter and nuclear reactions convert the former WD material into heavier elements, some of which escape in an accretion disk wind.

Much of the escaping material is made up of unburned matter entrained into the outflow, so C/O WD disruptions are unlikely to produce CaSTs due to the high O abundances in the ejecta \citep{zena19a,zena19b}.  However, He WD disruptions by NSs are a promising progenitor channel, as their ejecta are dominated by intermediate-mass elements, including Ca, $1-3\E{-3} \msol$ of $^{56}$Ni, and large amounts of unburned He (\citealt{marg16a}, although see \citealt{zena19a}, whose He WD + NS models did not give rise to mass ejection).

While He WD + NS binaries necessarily have relatively extreme mass ratios and have previously been assumed to undergo stable mass transfer, \cite{bobr17a}  challenged this assumption and suggested that systems with WD masses as low as $0.2 \msol$ still undergo unstable mass transfer and disruption of the WD.  This is due to accretion disk winds from the initial phase of super-Eddington mass transfer, which extract angular momentum and lead to merger.  Thus, most WD + NS binaries may lead to the disruption of the WD and subsequent ejection of thermonuclear ash.

He WD + NS binaries may have a similar velocity distribution to double NS binaries due to the natal kicks of the NSs.  The radial distribution of He WD + NS binaries at the time of interaction may be even broader than the distribution of  short gamma-ray bursts, which presumably arise from double NS binaries, because of the long lifetime of the main sequence star that eventually forms the He WD.  The NS's natal kick coupled with the long travel time prior to interaction could thus potentially explain CaSTs and their radial distribution, which is indeed more extended than the short gamma-ray burst distribution.

Unfortunately, the estimated rate of WD+NS mergers in the field is $\sim 3\E{-16} - 3\E{-15} {\rm \, yr^{-1} \,} M_\odot^{-1}$ for a $\unit[5]{Gyr}$-old population \citep{toon18b}.  This is $0.3-3\%$ of the Type Ia SN rate at the same age, well below the observed CaST rate.  Moreover, only a small fraction of these WD+NS systems will harbor a He WD, which we assume to be necessary to avoid large amounts of O in the ejecta.

However, as with the compact binaries discussed in the previous section, the rate of formation and interaction will be dramatically enhanced inside of GCs.  Again, further work is required, but we regard He WD disruptions by NSs as another viable CaST progenitor, bringing our number of candidate scenarios to three: He WD + NS binaries, He + O/Ne WD binaries, and low-mass He-burning star + WD binaries.


\section{Globular clusters as production sites of calcium-strong transients}
\label{sec:gc}

The radial distribution of CaSTs with respect to their host galaxies is perhaps their most unique characteristic.  As shown in Figure \ref{fig:gccdf}, the average offset for the gold and silver CaST samples is $39$ and $\unit[24]{kpc}$, respectively.  Several studies have suggested this offset could be due to hypervelocity ejection of a progenitor binary from a host galaxy, possibly due to an interaction with a supermassive black hole binary at the center of the galaxy (e.g., \citealt{lyma14a,lyma16a,fole15b,coug18a}; see also \citealt{wang19a}).  The most quantitative observational work on this is by \cite{fole15b}, who inferred velocity shifts from asymmetries in Ca nebular line profiles and used this link as evidence for hypervelocity ejection.

There are two main objections to the hypervelocity ejection scenario.  With regards to the velocities inferred from the asymmetric Ca nebular lines, \cite{mili17a} showed that, while the Ca nebular lines may show asymmetries, other nebular lines do not.  This suggests that the asymmetries are not due to velocity shifts but are instead inherent to the Ca lines, perhaps from preferential absorption of the redshifted component by the explosion ejecta.

Another issue arises from the CaST rates.  As we discuss in Section \ref{sec:rates}, the CaST rate is roughly 40\% of the local Type Ia SN rate, whereas \cite{coug18a} estimate a rate of double WD ejections by supermassive black holes, the most commonly invoked ejected progenitor binary, of only $0.001-0.1$ of the Type Ia SN rate.  Furthermore, the progenitors of CaSTs must also explode at a low rate within their host galaxies in order to reproduce their radial distribution.  This rules out typical double WD binaries, which merge within the field at a rate $\sim 6$ times higher than the Type Ia SN rate \citep{maoz18a}.  The necessary combination of a low rate within galaxies and a high rate outside of galaxies makes ejection by supermassive black holes implausible.


\subsection{The globular cluster radial distribution}

We now derive the GC galactocentric radial distribution in the local universe and show that it is consistent with the CaST radial distribution.  The CaST class may thus arise from rare binaries that do not interact often in the field but whose birth and interaction rates are enhanced by dynamical processes within GCs; we explore this further in Section \ref{sec:mechanism}.  We note that this work is complementary to, but distinct from, that of \cite{yuan13a}, who modeled the GC profiles for the host galaxies of the CaSTs known at the time and found them to be consistent with the CaST distribution (although they discarded a GC association due to the non-detections of GCs at the sites of CaSTs).  Here, we derive the theoretical radial distribution of all GCs in the local universe without restriction to just those in host galaxies of CaSTs and show that the GC distribution is consistent with the distribution of CaSTs.

In order to calculate the GC radial distribution function, we connect the properties of GCs to their host halos.  The total mass of GCs associated with a galaxy does not track the galaxy's stellar mass in a simple way.  However, the total mass in GCs, $M_{\rm GCs}$, does vary essentially linearly with the galaxy's halo mass, $M_{\rm halo}$, as  $M_{\rm GCs}/M_{\rm halo} \sim 3\E{-5}$ across five orders of magnitude, with no strong dependence on galaxy type (e.g., \citealt{blak97a,harr15a,chok19a}).  We can thus use the well-characterized halo mass function (HMF) at redshift zero to derive properties of GCs in the local universe.

The galactocentric half-mass, or effective, radii, $R_e$, of GC systems are, on average, several times larger than those of their host galaxies.  \cite{forb17a} compiled a sample of GC $R_e$ around early-type galaxies and demonstrated several correlations, including a dependence with halo mass that is consistent with a power-law slope,
\begin{align}
	R_e \sim \unit[22]{kpc} \, \left( \frac{M_{\rm halo}}{10^{13} \msol} \right)^{1/3} .
	\label{eqn:reff}
\end{align}
This slope matches theoretical expectations of the relationship between galaxy and halo sizes \citep{krav13a}.\footnote{\cite{huds18a} note that the removal of ultra-diffuse galaxies, whose halo masses are difficult to determine, from \cite{forb17a}'s sample changes the power-law slope to $R_e \propto M_{\rm halo}^{0.74}$, which is consistent with their own analysis of a sample that includes late-type galaxies as well.  We leave an in-depth exploration of the effects of different slopes and scatter in this and the other relations we use to future work.}

The radial dependence of a GC system's surface density with respect to its host galaxy is often fit by a S\'{e}rsic profile,
\begin{align}
	\sigma(R) = \sigma_e \exp \left\{ -b_n \left[ \left( \frac{R}{R_e} \right)^{1/n}-1 \right]  \right\} ,
\end{align}
where $R$ is the projected offset, $\sigma_e$ is the surface density at the effective radius, $R_e$, and $b_n = 1.9992n - 0.3271$.  The index $n$ varies depending on the GC system; we use representative values of $n=2$ and $n=4$ (a de Vaucouleurs profile) in the following analysis.  Multiplying the S\'{e}rsic profile by $2 \pi R dR$ and integrating yields the cumulative distribution function of projected offsets.

\begin{figure}
  \centering
  \includegraphics[width=\columnwidth]{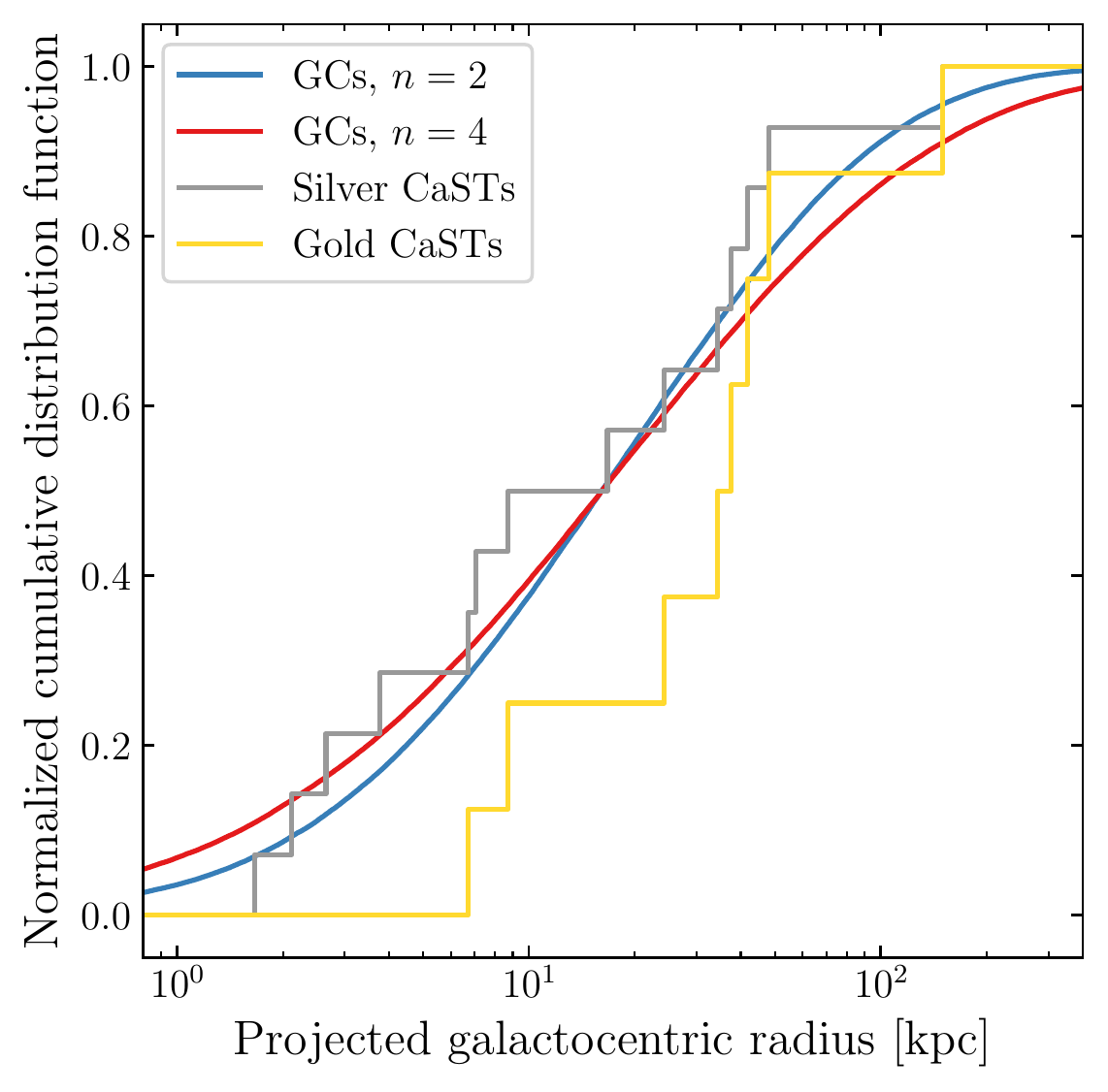}
  \caption{Projected galactocentric radial distribution functions of CaSTs and GCs.  The GC radial distributions assume S\'{e}rsic profiles with $n=2$ (blue line) and $n=4$ (red line).  The yellow and gray lines show the gold- and silver-sample radial distributions of CaSTs, respectively.}
  \label{fig:gccdf}
\end{figure}

We then integrate \cite{tink08a}'s parameterization of the HMF using \texttt{HMFcalc} \citep{murr13a} to obtain the mass-weighted halo cumulative distribution function.  We restrict the halos to those with masses between $10^{10}-10^{15} \msol$, where the lower bound is chosen because halos without GCs begin to appear $<10^{10} \msol$ \citep{forb18a}.  We sample from this mass-weighted halo CDF, which also corresponds to the mass-weighted GC distribution function, and relate the halo mass to the effective radius of the GC system using equation (\ref{eqn:reff}) to obtain the distribution function of projected offsets for GCs in the local universe.  This distribution is shown in Figure \ref{fig:gccdf} as blue and red lines for S\'{e}rsic indices of $n=2$ and $n=4$, respectively.

The gray and yellow lines show the observed projected offset distribution for the silver and gold samples.  The GC radial distributions do indeed match the silver sample's distribution very well.  Furthermore, as \cite{lunn17a} and \cite{froh18a} note, the Palomar Transient Factory, which discovered the majority of the CaSTs in the gold sample, is somewhat biased against recovering CaSTs closer to their host galaxies, so we view the gold sample's radial distribution as an upper limit on the physical radial distribution of CaSTs.


\subsection{Calcium-strong transient and globular cluster host galaxy properties}
\label{sec:gchosts}

It has been noted that CaSTs preferentially occur in galaxy groups and clusters \citep{fole15b,lunn17a}.  Moreover, in the gold sample, 88\% of the host galaxies are ellipticals, although this percentage decreases to 64\% for the silver sample.  In this section, we examine the host galaxy properties of GCs to determine if their galaxy types and richnesses with respect to clustering are consistent with those of CaSTs.

We again make use of the linear correlation between GC system mass and halo mass and then connect the halo mass to the galaxy stellar mass using the functional fit from \cite{behr10a}, which holds for both early- and late-type galaxies \citep{wech18a}.  We then use the compilation of \cite{blan09a} (their Fig.\ 4), which separates galaxies into types and clustering richness.  We find that, if a GC is selected at random in the local universe, its host galaxy will be a spiral 37\% and an elliptical 63\% of the time.  This fraction is remarkably consistent with the silver sample of CaSTs.

Furthermore, only 24\% of  galaxies selected this way are isolated.  However, only 17\% of GC hosts are in truly rich environments with at least 10 nearby neighbors, while 75\% of the gold sample hosts are in rich environments.  Data for the clustering of the silver sample's host galaxies does not exist in the literature, so we do not make a comparison.  Small numbers and selection bias may play a role in the richness of the gold sample hosts; a larger sample of CaSTs, which will soon be available from surveys such as the Zwicky Transient Facility and the Large Synoptic Survey Telescope, will better quantify the properties of their host galaxies and allow a more precise comparison to GC hosts.


\subsection{Dynamical formation and binary hardening within, and ejection from, globular clusters}
\label{sec:mechanism}

The overall correspondence of the GC and CaST radial distributions and their host galaxy properties described in the previous sections suggests a causal link: perhaps CaSTs are produced within GCs.  The extremely dense environments of GCs allow for the creation of stellar binaries via dynamical processes including tidal captures, three-body-induced captures, exchanges, and direct collisions with giant envelopes.  These yield theorized and observed enhanced formation rates, relative to the field, of stellar binaries and their end products, including cataclysmic variables, double WD binaries, low-mass X-ray binaries, ultra-compact X-ray binaries, and millisecond pulsars (for a review, see \citealt{bena13a}).  We do not conduct a quantitative estimate of the birth rates of possible CaST progenitor binaries formed in these ways, as this would require binary population synthesis calculations, including uncertain NS formation via WD collapse and electron-capture SNe, and $N$-body simulations that are beyond the scope of this work (see, e.g., \citealt{ivan06a,ivan08a} and \citealt{sams17a}).  Instead, we proceed under the assumption that CaST progenitors are formed dynamically in GCs as hard binaries at a rate larger than, or at least competitive with, that in the field and leave a detailed study of the birth rates of our proposed progenitor binaries in GCs to future work.

In addition to enhancing the formation of CaST progenitor binaries, GCs will also increase their rate of interaction via binary hardening, which decreases the binary separation for hard binaries on a timescale of
\begin{align}
	t_{\rm harden} & \sim  \,  5.4 {\rm \, Gyr} \left( \frac{\xi}{0.3} \right)^{-1} \frac{(M_1+M_2)M_\odot}{M_3 M_t} \nonumber \\
	\times & \left( \frac{\sigma}{10 \, {\rm km \, s^{-1}}} \right)\left( \frac{n}{10^7 \, {\rm pc^{-3}}} \right)^{-1} \left( \frac{a}{R_\odot} \right)^{-1} ,
\end{align}
where $\sigma$ is the GC velocity dispersion, $n$ is the GC stellar density (the fiducial value of $\unit[10^7]{pc^{-3}}$ assumes a core-collapsed GC), $a$ is the binary separation, and $\xi$ is a factor that parameterizes the average change in binding energy resulting from a flyby \citep{hegg75a,hill83a,spit87a}.  The binary component masses are $M_1$ and $M_2$, the mass of a typical GC star is $M_3$, and $M_t=M_1+M_2+M_3$.  For the He WD + NS, He WD + O/Ne WD, and low-mass He-burning star + WD progenitor binaries we are considering, $M_1 \simeq (0.3, 0.3, 0.5) \, M_\odot$ and $M_2 \simeq (1.4, 1.2, 0.7) \, M_\odot$, respectively, and we take the typical mass of a GC star to be $M_3 \simeq 0.7 \, M_\odot$.

Hard binaries will harden until either gravitational wave emission dominates the further evolution of the binary or subsequent hardening results in a recoil velocity that ejects the binary from the GC (for similar discussions of binary black hole and binary NS ejection, see, e.g., \citealt{rodr16a} and \citealt{andr19a}).  Equating the recoil velocity to the GC's escape velocity results in a critical separation below which further hardening results in ejection \citep{sigu93a}:
\begin{align}
	a_{\rm escape} & \sim  \, 23 \, R_\odot \left( \frac{\xi}{0.3} \right) \left( \frac{v_{\rm escape}}{50 \, {\rm km \, s^{-1}}} \right)^{-2} \nonumber \\
	\times & \frac{ M_1 M_2 M_3^2}{M_t(M_1+M_2)^2 M_\odot }  ,
\end{align}
where $v_{\rm escape}$ is the GC escape velocity.  The term involving masses ranges from $0.03-0.07$ for the progenitor binaries we are considering.

The ratio of this critical separation to the separation at which gravitational wave emission dominates is
\begin{align}
	\frac{a_{\rm escape}}{a_{\rm GWR}}  & \sim \, 11.2 \left( \frac{\xi}{0.3} \right)^{6/5} \left( \frac{n}{10^7 \, {\rm pc^{-3}}} \right)^{1/5} \left( \frac{v_{\rm esc}}{50 \, {\rm km \, s^{-1}}} \right)^{-2} \nonumber \\
	 \times & \left( \frac{\sigma}{10 \, {\rm km \, s^{-1}}} \right)^{-1/5} \left( \frac{M_1^4 M_2^4 M_3^{11} }{M_t^4(M_1+M_2)^{12} M_\odot^3 } \right)^{1/5} .
\end{align}
The mass term ranges from $0.03-0.09$, so $a_{\rm escape}$ varies from a few times smaller than $a_{\rm GWR}$ up to the same order of magnitude.  Thus, it is plausible that at least some of the CaST progenitor binaries we are considering will be ejected from the GC prior to interacting.  However, it is also possible that some of the binaries will still be in their birth GCs when the explosion occurs, especially for clusters with larger escape velocities.  Simulations coupling binary population synthesis and $N$-body dynamics are necessary to better quantify this question; for the present, we encourage continued deep searches for GCs at the sites of CaSTs.


\subsection{The calcium-strong transient rate and the implied production efficiency}
\label{sec:rates}

Integrating the mass-weighted GC CDF yields the total mass density of GCs in the local universe: $\rho_{\rm GCs} = 1.4\E{6} \msol {\rm \, Mpc^{-3}}$.  Recently, \cite{froh18a} calculated the volumetric CaST rate to be $ \unit[1.21^{+1.13}_{-0.39}\E{-5}]{yr^{-1}} {\rm \, Mpc^{-3}}$.  If CaSTs are produced by stars originating in GCs and have been occurring at a constant rate for a Hubble time, this implies an efficiency of
\begin{align}
	\eta \sim \frac{\textrm{1 CaST}}{8.4 \, M_\odot\textrm{ in GCs}} .
\end{align}
If instead these transients have a long delay between stellar birth and explosion (e.g., due to long main sequence lifetimes for stars that eventually become He WDs or low-mass He-burning stars and due to long binary hardening and gravitational inspiral timescales) and have only been occurring for the past $\unit[1]{Gyr}$, $\eta \sim 1/120 \, M_\odot$.  The required efficiency also decreases further if most GCs were more massive in the past (\citealt{grat12a} and references therein).

Integrating a Kroupa stellar initial mass function, assuming all stars between $1$ and $8 \msol$ have become WDs, and ignoring stars $>8 \msol$ (because many have become NSs with large natal kicks that ejected them from the GC or because they become stellar-mass black holes with negligible contribution to the total GC mass), we find $0.3$ WDs per $M_\odot$ in present-day GCs.  If CaSTs have been occurring at a constant rate for a Hubble time, this would imply that nearly half of all WDs would have to be involved in CaSTs.  If instead CaSTs have only been occurring for the last Gyr, 3\% of WDs in GCs have been involved in CaSTs.

One mitigating factor is the combined effect of relaxation, mass segregation, and the galaxy's tidal potential, which lead to a concentration of WDs in GC centers accompanied by preferential stripping of low-mass main sequence stars from the GC outskirts.  These effects boost the fraction of GC mass in WDs by a factor of $2-7$, depending on the age of the GC \citep{vesp97a,baum03a}, which significantly relaxes the required fraction of WDs that participate in CaSTs.  Still, even with these effects, the required efficiency remains quite high, especially since only some GCs have the high stellar densities required to harden binaries sufficiently within a Hubble time.  This issue becomes even starker if a He WD is required, as these are likely a small percentage of the total WD population.  The rate of formation of the proposed CaST progenitors within GCs thus represents an important constraint that should be tested with $N$-body and binary population synthesis calculations.

Similar consideration of the NS population within GCs suggests that He WD + NS binaries are not viable CaST progenitors.  \cite{ivan08a} find only one NS retained per $800-900 \, M_\odot$ in a present-day GC.  This fraction is significantly less than the required mass-specific CaST production efficiency, $\eta$, derived above.  For this channel to succeed, the NSs would have to be produced and retained in GCs at much higher rates than currently expected.


\subsection{Calcium-strong transients in nuclear star clusters}
\label{sec:nsc}

If high stellar densities and lifetimes long enough to form He WDs or low-mass He-burning stars are sufficient conditions for the production of CaSTs, nuclear star clusters (NSCs) should also give rise to CaSTs.  NSCs have stellar densities as high as $ 10^5 \, M_\odot \, {\rm pc^{-2}}$, and, while they do possess young stars, they also harbor a large population of old stars as well.  In fact, one NSC formation mechanism invokes the inspiral of GCs via dynamical friction into the galactic center where they are disrupted to form the NSC (see, e.g., \citealt{gned14a}).  The connection of gamma-ray emission in the Milky Way's galactic center to a postulated population of millisecond pulsars that may come from such disrupted GCs is one possible piece of evidence in support of this formation mechanism \citep{bran15a,frag18a}.

With their larger masses and similar radii, NSCs have escape velocities several times larger than those from GCs \citep{anto16a}, and so binary hardening recoil kicks may not be able to eject any CaST progenitors from NSCs.  Furthermore, given the large amount of mass near galactic centers, even CaST progenitors that are ejected from NSCs will not travel very far radially from their galactic nuclei.  Thus, CaST progenitors born in NSCs will explode inside or near their parent NSCs, leading to a sharp spike in the CaST galactocentric radial distribution centered on galactic nuclei.  As the masses in NSCs are roughly comparable to GC system masses \citep{sanc19a}, there should be a similar number of CaSTs centered on galactic nuclei as in the outskirts of galaxies if CaSTs are indeed formed by dynamical processes.  

No CaSTs have yet been observed near galactic nuclei; the closest in the silver (gold) sample is at an offset of $1.7$ $\unit[(6)]{kpc}$.  However, as quantified by \cite{froh18a}, CaSTs are very difficult to recover in the bright cores of galaxies, and so the lack of detection of any CaST in a galactic nucleus is not yet constraining.  Future observations with the ability to recover CaSTs in galaxy cores will provide an important test of CaST progenitor scenarios involving dynamical formation processes.


\section{The galactocentric radial distribution of old, metal-poor stars}
\label{sec:oldlowz}

\cite{yuan13a} and \cite{pere14a} have suggested that the large galactocentric offsets of CaSTs may be explained if they arise from an old, metal-poor stellar population, as such stars may preferentially reside in galaxy halos.  In this section, following the analysis of  \cite{yuan13a}, we examine the galactocentric radial distributions of stellar populations of various ages and metallicities using the results of TNG100, part of the IllustrisTNG cosmological galaxy formation simulation suite \citep{mari18a,naim18a,nels18a,nels19a,pill18a,spri18a}.  Although it would be preferable to carry out this analysis using observational data of stellar halos, it is not feasible to obtain accurate metallicities and ages in most cases.  We note that the overall stellar density profiles from IllustrisTNG are consistent with observed stacked galaxy data \citep{dsou14a}, giving confidence that the more detailed binning of IllustrisTNG data by age and metallicity accurately reflects reality.

\begin{figure}
  \centering
  \includegraphics[width=\columnwidth]{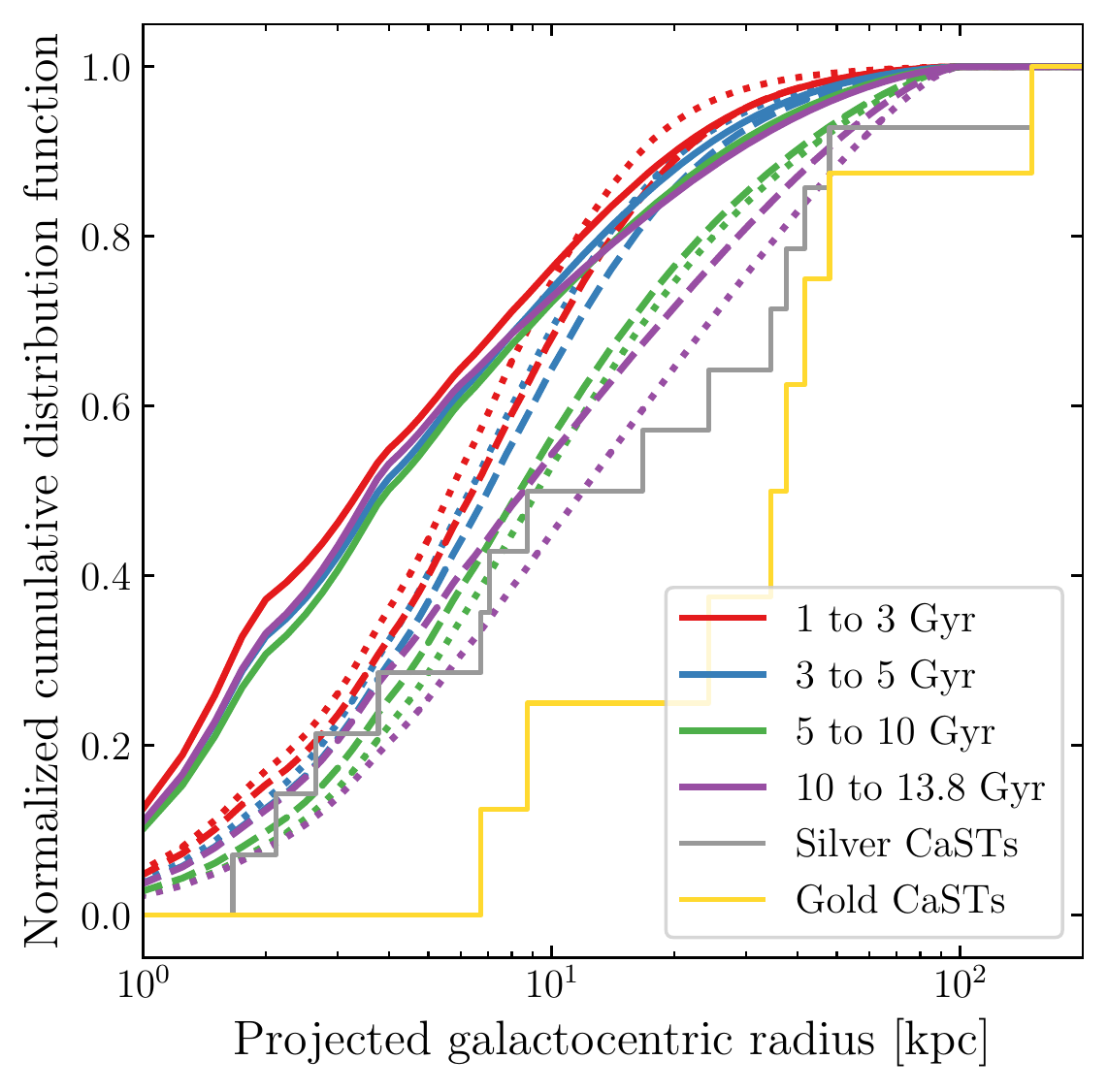}
  \caption{Projected galactocentric radial distribution functions of CaSTs and stellar populations of various ages and metallicities from TNG100.  Red, blue, green, and purple lines show distributions of stellar populations with ages from $1$ to $3$, $3$ to $5$, $5$ to $10$, and $ 10$ to $ \unit[13.8]{Gyr}$, and solid, dashed, and dotted lines show distributions of stars with metallicities of $Z/Z_\odot \geq 0.3$, $0.3 > Z/Z_\odot \geq 0.1$, and $Z/Z_\odot < 0.1$, respectively.  As in Figure \ref{fig:gccdf}, the yellow and gray lines show the gold- and silver-sample radial distributions of CaSTs.}
  \label{fig:stellarpopcdf}
\end{figure}

Figure \ref{fig:stellarpopcdf} shows the projected galactocentric radial distributions of the gold- and silver-sample CaSTs and of TNG100 stars in age bins of $1$ to $3$ (red), $3$ to $5$ (blue), $5$ to $10$ (green), and $10$ to $\unit[13.8]{Gyr}$ (purple lines) and metallicity bins of $Z/Z_\odot \geq 0.3$ (solid), $0.3 > Z/Z_\odot \geq 0.1$ (dashed), and $Z/Z_\odot < 0.1$ (dotted lines), where $Z_\odot$ is the solar metallicity.  Near-solar metallicity stars of all ages are inconsistent with the CaST distribution.  Intriguingly, the radial distribution of very old ($ > \unit[10]{Gyr}$), very metal-poor ($ <0.1 \, Z_\odot$) stars is consistent with the silver sample of CaSTs.  The distribution of somewhat younger ($ > \unit[5]{Gyr}$) and higher metallicity ($ < 0.3 \, Z_\odot$) stars is also marginally compatible with the silver sample.  These results are broadly consistent with those found by \cite{yuan13a}.

Using \cite{blan09a}'s compilation and performing a similar analysis to that in Section \ref{sec:gchosts}, we find that 71\% of the oldest, most metal-poor TNG100 stellar population is associated with an elliptical galaxy, 19\% of their host galaxies are isolated, and 26\% are in clustered environments.  These numbers only change slightly when considering the $> \unit[5]{Gyr}$, $Z/Z_\odot < 0.3$ population.  Similarly to GCs, these numbers provide a good match to the silver CaST sample's elliptical galaxy fraction but not to the gold sample's high percentage of hosts in clustered environments.

There is much more mass in old, metal-poor stars than in GCs, and so the required efficiency to produce CaSTs is a much more reasonable 1 CaST per $130 \, M_\odot$ if we only consider the oldest, lowest metallicity population and if CaSTs have occurred  at a constant rate for only the last $\unit[3.8]{Gyr}$.  If CaSTs instead have a delay time of $\unit[5]{Gyr}$ after a burst of star formation but can occur at metallicities up to $0.3 \, Z_\odot$, the required efficiency decreases even further to 1 CaST per $250 \, M_\odot$.

However, while old, low-metallicity stellar populations have a relatively large reservoir of mass and provide an interesting match to the CaST radial distribution and elliptical galaxy fraction, no obvious progenitor scenario presents itself.  Requiring an old stellar population is sensible, but it is not clear why the progenitors should be metal-poor.  For example, binaries involving a He WD or low-mass He-burning star will only be formed at relatively long delay times, but it is not obvious why these would require low metallicities to become CaSTs.  There is evidence that the binary fraction of solar-type stars is as much as twice as high at $Z = 0.1 \, Z_\odot$ as compared to solar metallicity \citep{moe19a}, but the main sequence mass of the primary star in any binary that gives rise to a CaST is likely more massive than a solar-type star.  In addition, this effect would not offset the much larger amount of mass in higher metallicity stars, which are not found at large radial offsets in the IllustrisTNG data.


\section{Conclusions}

In this work, we have examined the recently discovered class of enigmatic ``calcium-strong transients'' (CaSTs), comparing their observed properties to a wide variety of proposed progenitor models.  Their high Ca/O nebular line ratios, extended galactocentric radial distributions, and relatively high occurrence rate strongly constrain the allowed progenitor scenarios.  We use the Illustris TNG100 simulation data to derive the theoretical galactocentric radial distribution of stars of various ages and metallicities and find that old ($ > \unit[5]{Gyr}$), metal-poor ($ Z/Z_\odot < 0.3$) stars are consistent with the CaST radial distribution; however, no obvious progenitor scenario requiring such extreme stellar properties presents itself.

Additionally, we examine globular clusters (GCs) in the local universe and find that their radial distribution is also consistent with that of the CaSTs.  We propose a progenitor scenario in which one of three types of binary systems -- He + O/Ne WDs, He WDs + NSs, or low-mass He-burning stars + WDs, each of which is rare in the field -- is instead formed dynamically at an enhanced rate inside GCs.  Subsequent binary hardening both increases the interaction rate and leads to recoil kicks that may eventually eject the binaries from the GC, which explains why no GC has yet been discovered at the site of a CaST.  However, this explanation suffers from the relatively small amount of mass in GCs, which requires a high efficiency of CaST production per unit mass in GCs.  Such considerations appear to rule out He WDs + NSs as CaST progenitors due to the small number of NSs retained in GCs.

Future work is required to better quantify these proposed progenitors.  Multidimensional reactive hydrodynamic simulations of each of the models will yield more precise outputs with which to compare to actual observations.  Dynamical $N$-body calculations, including binary stellar evolution, will allow comparisons to the observed CaST rate as well as test the hypothesis that the progenitor binaries are ejected from their GCs prior to interacting.  At present, our estimates suggest that some of the CaSTs may remain inside their parent GCs at the time of explosion; we encourage continued follow-up of future CaSTs to search for host GCs.  We predict that if CaSTs are indeed dynamically formed in GCs, they should also explode in nuclear star clusters at a similar rate to the presently observed CaST rate; future observations with the power to distinguish faint CaSTs in bright galaxy cores will provide a crucial test.  A larger sample of CaSTs will also better constrain the observed radial distribution, which we expect to behave more like the silver sample of CaSTs than the gold.


\acknowledgments

K.J.S.\ thanks Mr.\ Woosnam for being his first mentor and a lifelong role model.  We acknowledge helpful discussions with Abbas Askar, Peter Behroozi, Nick Choksi, Kishalay De, Kareem El-Badry, Jim Fuller, Mirek Giersz, Mansi Kasliwal, Nathan Leigh, Alison Miller, Hagai Perets, Sterl Phinney, Josiah Schwab, and Tom Zick.  Support for this work was provided by NASA through the Astrophysics Theory Program (NNX17AG28G), by a Simons Investigator award from the Simons Foundation (E.Q.), and by the Gordon and Betty Moore Foundation through Grant GBMF5076.  E.Q.\ thanks the Princeton Astrophysical Sciences department and the theoretical astrophysics group and Moore Distinguished Scholar program at Caltech for their hospitality and support.


\software{\texttt{HMFcalc} \citep{murr13a}}




\begin{thebibliography}{}
\expandafter\ifx\csname natexlab\endcsname\relax\def\natexlab#1{#1}\fi
\providecommand{\url}[1]{\href{#1}{#1}}
\providecommand{\dodoi}[1]{doi:~\href{http://doi.org/#1}{\nolinkurl{#1}}}
\providecommand{\doeprint}[1]{\href{http://ascl.net/#1}{\nolinkurl{http://ascl.net/#1}}}
\providecommand{\doarXiv}[1]{\href{https://arxiv.org/abs/#1}{\nolinkurl{https://arxiv.org/abs/#1}}}

\bibitem[{{Andrews} \& {Mandel}(2019)}]{andr19a}
{Andrews}, J.~J., \& {Mandel}, I. 2019, \apjl, 880, L8,
  \dodoi{10.3847/2041-8213/ab2ed1}

\bibitem[{{Anninos} {et~al.}(2018){Anninos}, {Fragile}, {Olivier}, {Hoffman},
  {Mishra}, \& {Camarda}}]{anni18a}
{Anninos}, P., {Fragile}, P.~C., {Olivier}, S.~S., {et~al.} 2018, \apj, 865, 3,
  \dodoi{10.3847/1538-4357/aadad9}

\bibitem[{{Antonini} \& {Rasio}(2016)}]{anto16a}
{Antonini}, F., \& {Rasio}, F.~A. 2016, \apj, 831, 187,
  \dodoi{10.3847/0004-637X/831/2/187}

\bibitem[{{Baumgardt} \& {Makino}(2003)}]{baum03a}
{Baumgardt}, H., \& {Makino}, J. 2003, \mnras, 340, 227,
  \dodoi{10.1046/j.1365-8711.2003.06286.x}

\bibitem[{{Behroozi} {et~al.}(2010){Behroozi}, {Conroy}, \&
  {Wechsler}}]{behr10a}
{Behroozi}, P.~S., {Conroy}, C., \& {Wechsler}, R.~H. 2010, \apj, 717, 379,
  \dodoi{10.1088/0004-637X/717/1/379}

\bibitem[{{Benacquista} \& {Downing}(2013)}]{bena13a}
{Benacquista}, M.~J., \& {Downing}, J.~M.~B. 2013, Living Reviews in
  Relativity, 16, 4, \dodoi{10.12942/lrr-2013-4}

\bibitem[{{Bildsten} {et~al.}(2007){Bildsten}, {Shen}, {Weinberg}, \&
  {Nelemans}}]{bild07}
{Bildsten}, L., {Shen}, K.~J., {Weinberg}, N.~N., \& {Nelemans}, G. 2007,
  \apjl, 662, L95, \dodoi{10.1086/519489}

\bibitem[{{Blakeslee} {et~al.}(1997){Blakeslee}, {Tonry}, \&
  {Metzger}}]{blak97a}
{Blakeslee}, J.~P., {Tonry}, J.~L., \& {Metzger}, M.~R. 1997, \aj, 114, 482,
  \dodoi{10.1086/118488}

\bibitem[{{Blanton} \& {Moustakas}(2009)}]{blan09a}
{Blanton}, M.~R., \& {Moustakas}, J. 2009, \araa, 47, 159,
  \dodoi{10.1146/annurev-astro-082708-101734}

\bibitem[{{Bobrick} {et~al.}(2017){Bobrick}, {Davies}, \& {Church}}]{bobr17a}
{Bobrick}, A., {Davies}, M.~B., \& {Church}, R.~P. 2017, \mnras, 467, 3556,
  \dodoi{10.1093/mnras/stx312}

\bibitem[{{Branch} {et~al.}(2004){Branch}, {Baron}, {Thomas}, {Kasen}, {Li}, \&
  {Filippenko}}]{bran04a}
{Branch}, D., {Baron}, E., {Thomas}, R.~C., {et~al.} 2004, \pasp, 116, 903,
  \dodoi{10.1086/425081}

\bibitem[{{Brandt} \& {Kocsis}(2015)}]{bran15a}
{Brandt}, T.~D., \& {Kocsis}, B. 2015, \apj, 812, 15,
  \dodoi{10.1088/0004-637X/812/1/15}

\bibitem[{{Brown} {et~al.}(2016){Brown}, {Kilic}, {Kenyon}, \&
  {Gianninas}}]{brow16b}
{Brown}, W.~R., {Kilic}, M., {Kenyon}, S.~J., \& {Gianninas}, A. 2016, \apj,
  824, 46, \dodoi{10.3847/0004-637X/824/1/46}

\bibitem[{{Carter} {et~al.}(2013){Carter}, {Marsh}, {Steeghs}, {Groot},
  {Nelemans}, {Levitan}, {Rau}, {Copperwheat}, {Kupfer}, \&
  {Roelofs}}]{cart13a}
{Carter}, P.~J., {Marsh}, T.~R., {Steeghs}, D., {et~al.} 2013, \mnras, 429,
  2143, \dodoi{10.1093/mnras/sts485}

\bibitem[{{Chen} {et~al.}(2019){Chen}, {Dong}, {Stritzinger}, {Holmbo},
  {Strader}, {Kochanek}, {Peng}, {Benetti}, {Bersier}, {Brownsberger},
  {Buckley}, {Gromadzki}, {Moran}, {Pastorello}, {Aydi}, {Bose}, {Connor},
  {Elias-Rosa}, {French}, {Holoien}, {Mattila}, {Shappee}, {Stark}, \&
  {Swihart}}]{chen19a}
{Chen}, P., {Dong}, S., {Stritzinger}, M.~D., {et~al.} 2019, \apjl, submitted
  (arXiv:1905.02205)

\bibitem[{{Choksi} \& {Gnedin}(2019)}]{chok19a}
{Choksi}, N., \& {Gnedin}, O.~Y. 2019, \mnras, 488, 5409,
  \dodoi{10.1093/mnras/stz2097}

\bibitem[{{Chu} \& {Li}(2007)}]{chu07a}
{Chu}, J., \& {Li}, W. 2007, CBET, 1084, 1

\bibitem[{{Clayton}(1996)}]{clay96}
{Clayton}, G.~C. 1996, \pasp, 108, 225, \dodoi{10.1086/133715}

\bibitem[{{Coughlin} {et~al.}(2018){Coughlin}, {Darbha}, {Kasen}, \&
  {Quataert}}]{coug18a}
{Coughlin}, E.~R., {Darbha}, S., {Kasen}, D., \& {Quataert}, E. 2018, \apjl,
  863, L24, \dodoi{10.3847/2041-8213/aad7bd}

\bibitem[{{De} {et~al.}(2018{\natexlab{a}}){De}, {Kasliwal}, {Cantwell}, {Cao},
  {Cenko}, {Gal-Yam}, {Johansson}, {Kong}, {Kulkarni}, {Lunnan}, {Masci},
  {Matuszewski}, {Mooley}, {Neill}, {Nugent}, {Ofek}, {Perrott},
  {Rebbapragada}, {Rubin}, {O'Sullivan}, \& {Yaron}}]{de18a}
{De}, K., {Kasliwal}, M.~M., {Cantwell}, T., {et~al.} 2018{\natexlab{a}}, \apj,
  866, 72, \dodoi{10.3847/1538-4357/aadf8e}

\bibitem[{{De} {et~al.}(2018{\natexlab{b}}){De}, {Kasliwal}, {Ofek}, {Moriya},
  {Burke}, {Cao}, {Cenko}, {Doran}, {Duggan}, {Fender}, {Fransson}, {Gal-Yam},
  {Horesh}, {Kulkarni}, {Laher}, {Lunnan}, {Manulis}, {Masci}, {Mazzali},
  {Nugent}, {Perley}, {Petrushevska}, {Piro}, {Rumsey}, {Sollerman},
  {Sullivan}, \& {Taddia}}]{de18b}
{De}, K., {Kasliwal}, M.~M., {Ofek}, E.~O., {et~al.} 2018{\natexlab{b}},
  Science, 362, 201, \dodoi{10.1126/science.aas8693}

\bibitem[{{de Plaa} {et~al.}(2007){de Plaa}, {Werner}, {Bleeker}, {Vink},
  {Kaastra}, \& {M{\'e}ndez}}]{depl07a}
{de Plaa}, J., {Werner}, N., {Bleeker}, J.~A.~M., {et~al.} 2007, \aap, 465,
  345, \dodoi{10.1051/0004-6361:20066382}

\bibitem[{{Dessart} \& {Hillier}(2015)}]{dess15a}
{Dessart}, L., \& {Hillier}, D.~J. 2015, \mnras, 447, 1370,
  \dodoi{10.1093/mnras/stu2520}

\bibitem[{{D'Souza} {et~al.}(2014){D'Souza}, {Kauffman}, {Wang}, \&
  {Vegetti}}]{dsou14a}
{D'Souza}, R., {Kauffman}, G., {Wang}, J., \& {Vegetti}, S. 2014, \mnras, 443,
  1433, \dodoi{10.1093/mnras/stu1194}

\bibitem[{{Ergma} \& {Fedorova}(1990)}]{ergm90a}
{Ergma}, E.~V., \& {Fedorova}, A.~V. 1990, \apss, 163, 143,
  \dodoi{10.1007/BF00639983}

\bibitem[{{Fern{\'a}ndez} {et~al.}(2019){Fern{\'a}ndez}, {Margalit}, \&
  {Metzger}}]{fern19a}
{Fern{\'a}ndez}, R., {Margalit}, B., \& {Metzger}, B.~D. 2019, \mnras, 488,
  259, \dodoi{10.1093/mnras/stz1701}

\bibitem[{{Fern{\'a}ndez} \& {Metzger}(2013)}]{fern13a}
{Fern{\'a}ndez}, R., \& {Metzger}, B.~D. 2013, \apj, 763, 108,
  \dodoi{10.1088/0004-637X/763/2/108}

\bibitem[{{Filippenko}(1997)}]{fili97a}
{Filippenko}, A.~V. 1997, \araa, 35, 309,
  \dodoi{10.1146/annurev.astro.35.1.309}

\bibitem[{{Filippenko} {et~al.}(2003){Filippenko}, {Chornock}, {Swift},
  {Modjaz}, {Simcoe}, \& {Rauch}}]{fili03a}
{Filippenko}, A.~V., {Chornock}, R., {Swift}, B., {et~al.} 2003, \iaucirc,
  8159, 2

\bibitem[{{Fink} {et~al.}(2007){Fink}, {Hillebrandt}, \& {R{\"o}pke}}]{fhr07}
{Fink}, M., {Hillebrandt}, W., \& {R{\"o}pke}, F.~K. 2007, \aap, 476, 1133,
  \dodoi{10.1051/0004-6361:20078438}

\bibitem[{{Fink} {et~al.}(2010){Fink}, {R{\"o}pke}, {Hillebrandt},
  {Seitenzahl}, {Sim}, \& {Kromer}}]{fink10}
{Fink}, M., {R{\"o}pke}, F.~K., {Hillebrandt}, W., {et~al.} 2010, \aap, 514,
  A53, \dodoi{10.1051/0004-6361/200913892}

\bibitem[{{Fink} {et~al.}(2014){Fink}, {Kromer}, {Seitenzahl},
  {Ciaraldi-Schoolmann}, {R{\"o}pke}, {Sim}, {Pakmor}, {Ruiter}, \&
  {Hillebrandt}}]{fink14a}
{Fink}, M., {Kromer}, M., {Seitenzahl}, I.~R., {et~al.} 2014, \mnras, 438,
  1762, \dodoi{10.1093/mnras/stt2315}

\bibitem[{{Foley}(2015)}]{fole15b}
{Foley}, R.~J. 2015, \mnras, 452, 2463, \dodoi{10.1093/mnras/stv789}

\bibitem[{{Forbes}(2017)}]{forb17a}
{Forbes}, D.~A. 2017, \mnras, 472, L104, \dodoi{10.1093/mnrasl/slx148}

\bibitem[{{Forbes} {et~al.}(2018){Forbes}, {Read}, {Gieles}, \&
  {Collins}}]{forb18a}
{Forbes}, D.~A., {Read}, J.~I., {Gieles}, M., \& {Collins}, M.~L.~M. 2018,
  \mnras, 481, 5592, \dodoi{10.1093/mnras/sty2584}

\bibitem[{{Fragione} {et~al.}(2018){Fragione}, {Antonini}, \&
  {Gnedin}}]{frag18a}
{Fragione}, G., {Antonini}, F., \& {Gnedin}, O.~Y. 2018, \mnras, 475, 5313,
  \dodoi{10.1093/mnras/sty183}

\bibitem[{{Frohmaier} {et~al.}(2018){Frohmaier}, {Sullivan}, {Maguire}, \&
  {Nugent}}]{froh18a}
{Frohmaier}, C., {Sullivan}, M., {Maguire}, K., \& {Nugent}, P. 2018, \apj,
  858, 50, \dodoi{10.3847/1538-4357/aabc0b}

\bibitem[{{Geier} {et~al.}(2013){Geier}, {Marsh}, {Wang}, {Dunlap}, {Barlow},
  {Schaffenroth}, {Chen}, {Irrgang}, {Maxted}, {Ziegerer}, {Kupfer},
  {Miszalski}, {Heber}, {Han}, {Shporer}, {Telting}, {G{\"a}nsicke},
  {{\O}stensen}, {O'Toole}, \& {Napiwotzki}}]{geie13a}
{Geier}, S., {Marsh}, T.~R., {Wang}, B., {et~al.} 2013, \aap, 554, A54,
  \dodoi{10.1051/0004-6361/201321395}

\bibitem[{{Gnedin} {et~al.}(2014){Gnedin}, {Ostriker}, \& {Tremaine}}]{gned14a}
{Gnedin}, O.~Y., {Ostriker}, J.~P., \& {Tremaine}, S. 2014, \apj, 785, 71,
  \dodoi{10.1088/0004-637X/785/1/71}

\bibitem[{{Gratton} {et~al.}(2012){Gratton}, {Carretta}, \&
  {Bragaglia}}]{grat12a}
{Gratton}, R.~G., {Carretta}, E., \& {Bragaglia}, A. 2012, \aapr, 20, 50,
  \dodoi{10.1007/s00159-012-0050-3}

\bibitem[{{Guillochon} {et~al.}(2010){Guillochon}, {Dan}, {Ramirez-Ruiz}, \&
  {Rosswog}}]{guil10}
{Guillochon}, J., {Dan}, M., {Ramirez-Ruiz}, E., \& {Rosswog}, S. 2010, \apjl,
  709, L64, \dodoi{10.1088/2041-8205/709/1/L64}

\bibitem[{{Harris} {et~al.}(2015){Harris}, {Harris}, \& {Hudson}}]{harr15a}
{Harris}, W.~E., {Harris}, G.~L., \& {Hudson}, M.~J. 2015, \apj, 806, 36,
  \dodoi{10.1088/0004-637X/806/1/36}

\bibitem[{{Heggie}(1975)}]{hegg75a}
{Heggie}, D.~C. 1975, \mnras, 173, 729, \dodoi{10.1093/mnras/173.3.729}

\bibitem[{{Hills}(1983)}]{hill83a}
{Hills}, J.~G. 1983, \aj, 88, 1269, \dodoi{10.1086/113418}

\bibitem[{{Hudson} \& {Robison}(2018)}]{huds18a}
{Hudson}, M.~J., \& {Robison}, B. 2018, \mnras, 477, 3869,
  \dodoi{10.1093/mnras/sty844}

\bibitem[{{Ivanova} {et~al.}(2008){Ivanova}, {Heinke}, {Rasio}, {Belczynski},
  \& {Fregeau}}]{ivan08a}
{Ivanova}, N., {Heinke}, C.~O., {Rasio}, F.~A., {Belczynski}, K., \& {Fregeau},
  J.~M. 2008, \mnras, 386, 553, \dodoi{10.1111/j.1365-2966.2008.13064.x}

\bibitem[{{Ivanova} {et~al.}(2006){Ivanova}, {Heinke}, {Rasio}, {Taam},
  {Belczynski}, \& {Fregeau}}]{ivan06a}
{Ivanova}, N., {Heinke}, C.~O., {Rasio}, F.~A., {et~al.} 2006, \mnras, 372,
  1043, \dodoi{10.1111/j.1365-2966.2006.10876.x}

\bibitem[{{Kasliwal} {et~al.}(2012){Kasliwal}, {Kulkarni}, {Gal-Yam}, {Nugent},
  {Sullivan}, {Bildsten}, {Yaron}, {Perets}, {Arcavi}, {Ben-Ami}, {Bhalerao},
  {Bloom}, {Cenko}, {Filippenko}, {Frail}, {Ganeshalingam}, {Horesh}, {Howell},
  {Law}, {Leonard}, {Li}, {Ofek}, {Polishook}, {Poznanski}, {Quimby},
  {Silverman}, {Sternberg}, \& {Xu}}]{kasl12}
{Kasliwal}, M.~M., {Kulkarni}, S.~R., {Gal-Yam}, A., {et~al.} 2012, \apj, 755,
  161, \dodoi{10.1088/0004-637X/755/2/161}

\bibitem[{{Kawabata} {et~al.}(2010){Kawabata}, {Maeda}, {Nomoto},
  {Taubenberger}, {Tanaka}, {Deng}, {Pian}, {Hattori}, \& {Itagaki}}]{kawa10a}
{Kawabata}, K.~S., {Maeda}, K., {Nomoto}, K., {et~al.} 2010, \nat, 465, 326,
  \dodoi{10.1038/nature09055}

\bibitem[{{Kawana} {et~al.}(2018){Kawana}, {Tanikawa}, \& {Yoshida}}]{kawa18a}
{Kawana}, K., {Tanikawa}, A., \& {Yoshida}, N. 2018, \mnras, 477, 3449,
  \dodoi{10.1093/mnras/sty842}

\bibitem[{{Khokhlov}(1991)}]{khok91a}
{Khokhlov}, A.~M. 1991, \aap, 245, 114

\bibitem[{{Kravtsov}(2013)}]{krav13a}
{Kravtsov}, A.~V. 2013, \apjl, 764, L31, \dodoi{10.1088/2041-8205/764/2/L31}

\bibitem[{{Long} {et~al.}(2014){Long}, {Jordan}, {van Rossum}, {Diemer},
  {Graziani}, {Kessler}, {Meyer}, {Rich}, \& {Lamb}}]{long14a}
{Long}, M., {Jordan}, IV, G.~C., {van Rossum}, D.~R., {et~al.} 2014, \apj, 789,
  103, \dodoi{10.1088/0004-637X/789/2/103}

\bibitem[{{Lunnan} {et~al.}(2017){Lunnan}, {Kasliwal}, {Cao}, {Hangard},
  {Yaron}, {Parrent}, {McCully}, {Gal-Yam}, {Mulchaey}, {Ben-Ami},
  {Filippenko}, {Fremling}, {Fruchter}, {Howell}, {Koda}, {Kupfer}, {Kulkarni},
  {Laher}, {Masci}, {Nugent}, {Ofek}, {Yagi}, \& {Yan}}]{lunn17a}
{Lunnan}, R., {Kasliwal}, M.~M., {Cao}, Y., {et~al.} 2017, \apj, 836, 60,
  \dodoi{10.3847/1538-4357/836/1/60}

\bibitem[{{Lyman} {et~al.}(2016{\natexlab{a}}){Lyman}, {Bersier}, {James},
  {Mazzali}, {Eldridge}, {Fraser}, \& {Pian}}]{lyma16b}
{Lyman}, J.~D., {Bersier}, D., {James}, P.~A., {et~al.} 2016{\natexlab{a}},
  \mnras, 457, 328, \dodoi{10.1093/mnras/stv2983}

\bibitem[{{Lyman} {et~al.}(2013){Lyman}, {James}, {Perets}, {Anderson},
  {Gal-Yam}, {Mazzali}, \& {Percival}}]{lyma13a}
{Lyman}, J.~D., {James}, P.~A., {Perets}, H.~B., {et~al.} 2013, \mnras, 434,
  527, \dodoi{10.1093/mnras/stt1038}

\bibitem[{{Lyman} {et~al.}(2014){Lyman}, {Levan}, {Church}, {Davies}, \&
  {Tanvir}}]{lyma14a}
{Lyman}, J.~D., {Levan}, A.~J., {Church}, R.~P., {Davies}, M.~B., \& {Tanvir},
  N.~R. 2014, \mnras, 444, 2157, \dodoi{10.1093/mnras/stu1574}

\bibitem[{{Lyman} {et~al.}(2016{\natexlab{b}}){Lyman}, {Levan}, {James},
  {Angus}, {Church}, {Davies}, \& {Tanvir}}]{lyma16a}
{Lyman}, J.~D., {Levan}, A.~J., {James}, P.~A., {et~al.} 2016{\natexlab{b}},
  \mnras, 458, 1768, \dodoi{10.1093/mnras/stw477}

\bibitem[{{MacLeod} {et~al.}(2016){MacLeod}, {Guillochon}, {Ramirez-Ruiz},
  {Kasen}, \& {Rosswog}}]{macl16a}
{MacLeod}, M., {Guillochon}, J., {Ramirez-Ruiz}, E., {Kasen}, D., \& {Rosswog},
  S. 2016, \apj, 819, 3, \dodoi{10.3847/0004-637X/819/1/3}

\bibitem[{{Maoz} {et~al.}(2018){Maoz}, {Hallakoun}, \& {Badenes}}]{maoz18a}
{Maoz}, D., {Hallakoun}, N., \& {Badenes}, C. 2018, \mnras, 476, 2584,
  \dodoi{10.1093/mnras/sty339}

\bibitem[{{Margalit} \& {Metzger}(2016)}]{marg16a}
{Margalit}, B., \& {Metzger}, B.~D. 2016, \mnras, 461, 1154,
  \dodoi{10.1093/mnras/stw1410}

\bibitem[{{Marinacci} {et~al.}(2018){Marinacci}, {Vogelsberger}, {Pakmor},
  {Torrey}, {Springel}, {Hernquist}, {Nelson}, {Weinberger}, {Pillepich},
  {Naiman}, \& {Genel}}]{mari18a}
{Marinacci}, F., {Vogelsberger}, M., {Pakmor}, R., {et~al.} 2018, \mnras, 480,
  5113, \dodoi{10.1093/mnras/sty2206}

\bibitem[{{Meng} \& {Han}(2015)}]{meng15a}
{Meng}, X., \& {Han}, Z. 2015, \aap, 573, A57,
  \dodoi{10.1051/0004-6361/201424562}

\bibitem[{{Mernier} {et~al.}(2016){Mernier}, {de Plaa}, {Pinto}, {Kaastra},
  {Kosec}, {Zhang}, {Mao}, {Werner}, {Pols}, \& {Vink}}]{mern16a}
{Mernier}, F., {de Plaa}, J., {Pinto}, C., {et~al.} 2016, \aap, 595, A126,
  \dodoi{10.1051/0004-6361/201628765}

\bibitem[{{Metzger}(2012)}]{metz12a}
{Metzger}, B.~D. 2012, \mnras, 419, 827,
  \dodoi{10.1111/j.1365-2966.2011.19747.x}

\bibitem[{{Milisavljevic} {et~al.}(2017){Milisavljevic}, {Patnaude}, {Raymond},
  {Drout}, {Margutti}, {Kamble}, {Chornock}, {Guillochon}, {Sanders},
  {Parrent}, {Lovisari}, {Chilingarian}, {Challis}, {Kirshner}, {Penny},
  {Itagaki}, {Eldridge}, \& {Moriya}}]{mili17a}
{Milisavljevic}, D., {Patnaude}, D.~J., {Raymond}, J.~C., {et~al.} 2017, \apj,
  846, 50, \dodoi{10.3847/1538-4357/aa7d9f}

\bibitem[{{Moe} {et~al.}(2019){Moe}, {Kratter}, \& {Badenes}}]{moe19a}
{Moe}, M., {Kratter}, K.~M., \& {Badenes}, C. 2019, \apj, 875, 61,
  \dodoi{10.3847/1538-4357/ab0d88}

\bibitem[{{Moriya} {et~al.}(2017){Moriya}, {Mazzali}, {Tominaga}, {Hachinger},
  {Blinnikov}, {Tauris}, {Takahashi}, {Tanaka}, {Langer}, \&
  {Podsiadlowski}}]{mori17a}
{Moriya}, T.~J., {Mazzali}, P.~A., {Tominaga}, N., {et~al.} 2017, \mnras, 466,
  2085, \dodoi{10.1093/mnras/stw3225}

\bibitem[{{Mulchaey} {et~al.}(2014){Mulchaey}, {Kasliwal}, \&
  {Kollmeier}}]{mulc14a}
{Mulchaey}, J.~S., {Kasliwal}, M.~M., \& {Kollmeier}, J.~A. 2014, \apjl, 780,
  L34, \dodoi{10.1088/2041-8205/780/2/L34}

\bibitem[{{Murray} {et~al.}(2013){Murray}, {Power}, \& {Robotham}}]{murr13a}
{Murray}, S.~G., {Power}, C., \& {Robotham}, A.~S.~G. 2013, Astronomy and
  Computing, 3, 23, \dodoi{10.1016/j.ascom.2013.11.001}

\bibitem[{{Naiman} {et~al.}(2018){Naiman}, {Pillepich}, {Springel},
  {Ramirez-Ruiz}, {Torrey}, {Vogelsberger}, {Pakmor}, {Nelson}, {Marinacci},
  {Hernquist}, {Weinberger}, \& {Genel}}]{naim18a}
{Naiman}, J.~P., {Pillepich}, A., {Springel}, V., {et~al.} 2018, \mnras, 477,
  1206, \dodoi{10.1093/mnras/sty618}

\bibitem[{{Nelemans} {et~al.}(2001){Nelemans}, {Portegies Zwart}, {Verbunt}, \&
  {Yungelson}}]{nele01b}
{Nelemans}, G., {Portegies Zwart}, S.~F., {Verbunt}, F., \& {Yungelson}, L.~R.
  2001, \aap, 368, 939, \dodoi{10.1051/0004-6361:20010049}

\bibitem[{{Nelson} {et~al.}(2018){Nelson}, {Pillepich}, {Springel},
  {Weinberger}, {Hernquist}, {Pakmor}, {Genel}, {Torrey}, {Vogelsberger},
  {Kauffmann}, {Marinacci}, \& {Naiman}}]{nels18a}
{Nelson}, D., {Pillepich}, A., {Springel}, V., {et~al.} 2018, \mnras, 475, 624,
  \dodoi{10.1093/mnras/stx3040}

\bibitem[{{Nelson} {et~al.}(2019){Nelson}, {Springel}, {Pillepich},
  {Rodriguez-Gomez}, {Torrey}, {Genel}, {Vogelsberger}, {Pakmor}, {Marinacci},
  {Weinberger}, {Kelley}, {Lovell}, {Diemer}, \& {Hernquist}}]{nels19a}
{Nelson}, D., {Springel}, V., {Pillepich}, A., {et~al.} 2019, Computational
  Astrophysics and Cosmology, 6, 2, \dodoi{10.1186/s40668-019-0028-x}

\bibitem[{{Nomoto}(1982)}]{nomo82b}
{Nomoto}, K. 1982, \apj, 257, 780, \dodoi{10.1086/160031}

\bibitem[{{Pakmor} {et~al.}(2013){Pakmor}, {Kromer}, {Taubenberger}, \&
  {Springel}}]{pakm13a}
{Pakmor}, R., {Kromer}, M., {Taubenberger}, S., \& {Springel}, V. 2013, \apjl,
  770, L8, \dodoi{10.1088/2041-8205/770/1/L8}

\bibitem[{{Paschalidis} {et~al.}(2011){Paschalidis}, {Liu}, {Etienne}, \&
  {Shapiro}}]{pasc11a}
{Paschalidis}, V., {Liu}, Y.~T., {Etienne}, Z., \& {Shapiro}, S.~L. 2011, \prd,
  84, 104032, \dodoi{10.1103/PhysRevD.84.104032}

\bibitem[{{Paschalidis} {et~al.}(2009){Paschalidis}, {MacLeod}, {Baumgarte}, \&
  {Shapiro}}]{pasc09a}
{Paschalidis}, V., {MacLeod}, M., {Baumgarte}, T.~W., \& {Shapiro}, S.~L. 2009,
  \prd, 80, 024006, \dodoi{10.1103/PhysRevD.80.024006}

\bibitem[{{Perets}(2014)}]{pere14a}
{Perets}, H.~B. 2014, arXiv e-prints, arXiv:1407.2254.
\newblock \doarXiv{1407.2254}

\bibitem[{{Perets} {et~al.}(2011){Perets}, {Gal-yam}, {Crockett}, {Anderson},
  {James}, {Sullivan}, {Neill}, \& {Leonard}}]{pere11}
{Perets}, H.~B., {Gal-yam}, A., {Crockett}, R.~M., {et~al.} 2011, \apjl, 728,
  L36+, \dodoi{10.1088/2041-8205/728/2/L36}

\bibitem[{{Perets} {et~al.}(2010){Perets}, {Gal-Yam}, {Mazzali}, {Arnett},
  {Kagan}, {Filippenko}, {Li}, {Arcavi}, {Cenko}, {Fox}, {Leonard}, {Moon},
  {Sand}, {Soderberg}, {Anderson}, {James}, {Foley}, {Ganeshalingam}, {Ofek},
  {Bildsten}, {Nelemans}, {Shen}, {Weinberg}, {Metzger}, {Piro}, {Quataert},
  {Kiewe}, \& {Poznanski}}]{pere10}
{Perets}, H.~B., {Gal-Yam}, A., {Mazzali}, P.~A., {et~al.} 2010, \nat, 465,
  322, \dodoi{10.1038/nature09056}

\bibitem[{{Pillepich} {et~al.}(2018){Pillepich}, {Nelson}, {Hernquist},
  {Springel}, {Pakmor}, {Torrey}, {Weinberger}, {Genel}, {Naiman}, {Marinacci},
  \& {Vogelsberger}}]{pill18a}
{Pillepich}, A., {Nelson}, D., {Hernquist}, L., {et~al.} 2018, \mnras, 475,
  648, \dodoi{10.1093/mnras/stx3112}

\bibitem[{{Polin} {et~al.}(2019){Polin}, {Nugent}, \& {Kasen}}]{poli19a}
{Polin}, A., {Nugent}, P., \& {Kasen}, D. 2019, \apj, 873, 84,
  \dodoi{10.3847/1538-4357/aafb6a}

\bibitem[{{Prentice} {et~al.}(2019){Prentice}, {Maguire}, {Fl{\"o}rs},
  {Taubenberger}, {Inserra}, {Frohmaier}, {Chen}, {Anderson}, {Ashall},
  {Clark}, {Fraser}, {Galbany}, {Gal-Yam}, {Gromadzki}, {Guti{\'e}rrez},
  {James}, {Jonker}, {Kankare}, {Leloudas}, {Magee}, {Mazzali}, {Nicholl},
  {Pursiainen}, {Skillen}, {Smartt}, {Smith}, {Vogl}, \& {Young}}]{pren19a}
{Prentice}, S.~J., {Maguire}, K., {Fl{\"o}rs}, A., {et~al.} 2019, \aap,
  submitted (arXiv:1909.05567)

\bibitem[{{Puckett} \& {Dowdle}(2000)}]{puck00a}
{Puckett}, T., \& {Dowdle}, G. 2000, \iaucirc, 7507, 2

\bibitem[{{Rodriguez} {et~al.}(2016){Rodriguez}, {Chatterjee}, \&
  {Rasio}}]{rodr16a}
{Rodriguez}, C.~L., {Chatterjee}, S., \& {Rasio}, F.~A. 2016, \prd, 93, 084029,
  \dodoi{10.1103/PhysRevD.93.084029}

\bibitem[{{Rosswog} {et~al.}(2008){Rosswog}, {Ramirez-Ruiz}, \&
  {Hix}}]{ross08a}
{Rosswog}, S., {Ramirez-Ruiz}, E., \& {Hix}, W.~R. 2008, \apj, 679, 1385,
  \dodoi{10.1086/528738}

\bibitem[{{Rosswog} {et~al.}(2009){Rosswog}, {Ramirez-Ruiz}, \&
  {Hix}}]{ross09b}
---. 2009, \apj, 695, 404, \dodoi{10.1088/0004-637X/695/1/404}

\bibitem[{{Samsing} {et~al.}(2017){Samsing}, {MacLeod}, \&
  {Ramirez-Ruiz}}]{sams17a}
{Samsing}, J., {MacLeod}, M., \& {Ramirez-Ruiz}, E. 2017, \apj, 846, 36,
  \dodoi{10.3847/1538-4357/aa7e32}

\bibitem[{{S{\'a}nchez-Janssen} {et~al.}(2019){S{\'a}nchez-Janssen},
  {C{\^o}t{\'e}}, {Ferrarese}, {Peng}, {Roediger}, {Blakeslee}, {Emsellem},
  {Puzia}, {Spengler}, {Taylor}, {{\'A}lamo-Mart{\'\i}nez}, {Boselli},
  {Cantiello}, {Cuillandre}, {Duc}, {Durrell}, {Gwyn}, {MacArthur},
  {Lan{\c{c}}on}, {Lim}, {Liu}, {Mei}, {Miller}, {Mu{\~n}oz}, {Mihos},
  {Paudel}, {Powalka}, \& {Toloba}}]{sanc19a}
{S{\'a}nchez-Janssen}, R., {C{\^o}t{\'e}}, P., {Ferrarese}, L., {et~al.} 2019,
  \apj, 878, 18, \dodoi{10.3847/1538-4357/aaf4fd}

\bibitem[{{Sell} {et~al.}(2018){Sell}, {Arur}, {Maccarone}, {Kotak}, {Knigge},
  {Sand}, \& {Valenti}}]{sell18a}
{Sell}, P.~H., {Arur}, K., {Maccarone}, T.~J., {et~al.} 2018, \mnras, 475,
  L111, \dodoi{10.1093/mnrasl/sly011}

\bibitem[{{Sell} {et~al.}(2015){Sell}, {Maccarone}, {Kotak}, {Knigge}, \&
  {Sand}}]{sell15a}
{Sell}, P.~H., {Maccarone}, T.~J., {Kotak}, R., {Knigge}, C., \& {Sand}, D.~J.
  2015, \mnras, 450, 4198, \dodoi{10.1093/mnras/stv902}

\bibitem[{{Shen}(2015)}]{shen15a}
{Shen}, K.~J. 2015, \apjl, 805, L6, \dodoi{10.1088/2041-8205/805/1/L6}

\bibitem[{{Shen} \& {Bildsten}(2009)}]{sb09b}
{Shen}, K.~J., \& {Bildsten}, L. 2009, \apj, 699, 1365,
  \dodoi{10.1088/0004-637X/699/2/1365}

\bibitem[{{Shen} \& {Bildsten}(2014)}]{shen14a}
---. 2014, \apj, 785, 61, \dodoi{10.1088/0004-637X/785/1/61}

\bibitem[{{Shen} {et~al.}(2018{\natexlab{a}}){Shen}, {Kasen}, {Miles}, \&
  {Townsley}}]{shen18a}
{Shen}, K.~J., {Kasen}, D., {Miles}, B.~J., \& {Townsley}, D.~M.
  2018{\natexlab{a}}, \apj, 854, 52, \dodoi{10.3847/1538-4357/aaa8de}

\bibitem[{{Shen} {et~al.}(2010){Shen}, {Kasen}, {Weinberg}, {Bildsten}, \&
  {Scannapieco}}]{shen10}
{Shen}, K.~J., {Kasen}, D., {Weinberg}, N.~N., {Bildsten}, L., \&
  {Scannapieco}, E. 2010, \apj, 715, 767, \dodoi{10.1088/0004-637X/715/2/767}

\bibitem[{{Shen} {et~al.}(2018{\natexlab{b}}){Shen}, {Boubert}, {G{\"a}nsicke},
  {Jha}, {Andrews}, {Chomiuk}, {Foley}, {Fraser}, {Gromadzki}, {Guillochon},
  {Kotze}, {Maguire}, {Siebert}, {Smith}, {Strader}, {Badenes}, {Kerzendorf},
  {Koester}, {Kromer}, {Miles}, {Pakmor}, {Schwab}, {Toloza}, {Toonen},
  {Townsley}, \& {Williams}}]{shen18b}
{Shen}, K.~J., {Boubert}, D., {G{\"a}nsicke}, B.~T., {et~al.}
  2018{\natexlab{b}}, \apj, 865, 15, \dodoi{10.3847/1538-4357/aad55b}

\bibitem[{{Sigurdsson} \& {Phinney}(1993)}]{sigu93a}
{Sigurdsson}, S., \& {Phinney}, E.~S. 1993, \apj, 415, 631,
  \dodoi{10.1086/173190}

\bibitem[{{Sim} {et~al.}(2012){Sim}, {Fink}, {Kromer}, {R{\"o}pke}, {Ruiter},
  \& {Hillebrandt}}]{sim12}
{Sim}, S.~A., {Fink}, M., {Kromer}, M., {et~al.} 2012, \mnras, 420, 3003,
  \dodoi{10.1111/j.1365-2966.2011.20162.x}

\bibitem[{{Spitzer}(1987)}]{spit87a}
{Spitzer}, L. 1987, {Dynamical evolution of globular clusters} (Princeton:
  Princeton University Press)

\bibitem[{{Springel} {et~al.}(2018){Springel}, {Pakmor}, {Pillepich},
  {Weinberger}, {Nelson}, {Hernquist}, {Vogelsberger}, {Genel}, {Torrey},
  {Marinacci}, \& {Naiman}}]{spri18a}
{Springel}, V., {Pakmor}, R., {Pillepich}, A., {et~al.} 2018, \mnras, 475, 676,
  \dodoi{10.1093/mnras/stx3304}

\bibitem[{{Sullivan} {et~al.}(2011){Sullivan}, {Kasliwal}, {Nugent}, {Howell},
  {Thomas}, {Ofek}, {Arcavi}, {Blake}, {Cooke}, {Gal-Yam}, {Hook}, {Mazzali},
  {Podsiadlowski}, {Quimby}, {Bildsten}, {Bloom}, {Cenko}, {Kulkarni}, {Law},
  \& {Poznanski}}]{sull11}
{Sullivan}, M., {Kasliwal}, M.~M., {Nugent}, P.~E., {et~al.} 2011, \apj, 732,
  118, \dodoi{10.1088/0004-637X/732/2/118}

\bibitem[{{Tanikawa}(2018)}]{tani18a}
{Tanikawa}, A. 2018, \apj, 858, 26, \dodoi{10.3847/1538-4357/aaba79}

\bibitem[{{Taubenberger}(2017)}]{taub17a}
{Taubenberger}, S. 2017, {in Handbook of Supernovae, ed. A.~W. Alsabti \& P.
  Murdin} (New York: Springer), 317, \dodoi{10.1007/978-3-319-21846-5_37}

\bibitem[{{Tauris} {et~al.}(2013){Tauris}, {Langer}, {Moriya}, {Podsiadlowski},
  {Yoon}, \& {Blinnikov}}]{taur13a}
{Tauris}, T.~M., {Langer}, N., {Moriya}, T.~J., {et~al.} 2013, \apjl, 778, L23,
  \dodoi{10.1088/2041-8205/778/2/L23}

\bibitem[{{Tinker} {et~al.}(2008){Tinker}, {Kravtsov}, {Klypin}, {Abazajian},
  {Warren}, {Yepes}, {Gottl{\"o}ber}, \& {Holz}}]{tink08a}
{Tinker}, J., {Kravtsov}, A.~V., {Klypin}, A., {et~al.} 2008, \apj, 688, 709,
  \dodoi{10.1086/591439}

\bibitem[{{Tonry} {et~al.}(2016){Tonry}, {Denneau}, {Stalder}, {Heinze},
  {Sherstyuk}, {Rest}, {Smith}, \& {Smartt}}]{tonr16a}
{Tonry}, J., {Denneau}, L., {Stalder}, B., {et~al.} 2016, ATel, 9685, 1

\bibitem[{{Toonen} {et~al.}(2018){Toonen}, {Perets}, {Igoshev}, {Michaely}, \&
  {Zenati}}]{toon18b}
{Toonen}, S., {Perets}, H.~B., {Igoshev}, A.~P., {Michaely}, E., \& {Zenati},
  Y. 2018, \aap, 619, A53, \dodoi{10.1051/0004-6361/201833164}

\bibitem[{{Townsley} {et~al.}(2019){Townsley}, {Miles}, {Shen}, \&
  {Kasen}}]{town19a}
{Townsley}, D.~M., {Miles}, B.~J., {Shen}, K.~J., \& {Kasen}, D. 2019, \apjl,
  878, L38, \dodoi{10.3847/2041-8213/ab27cd}

\bibitem[{{Valenti} {et~al.}(2014){Valenti}, {Yuan}, {Taubenberger}, {Maguire},
  {Pastorello}, {Benetti}, {Smartt}, {Cappellaro}, {Howell}, {Bildsten},
  {Moore}, {Stritzinger}, {Anderson}, {Benitez-Herrera}, {Bufano},
  {Gonzalez-Gaitan}, {McCrum}, {Pignata}, {Fraser}, {Gal-Yam}, {Le Guillou},
  {Inserra}, {Reichart}, {Scalzo}, {Sullivan}, {Yaron}, \& {Young}}]{vale14a}
{Valenti}, S., {Yuan}, F., {Taubenberger}, S., {et~al.} 2014, \mnras, 437,
  1519, \dodoi{10.1093/mnras/stt1983}

\bibitem[{{Vesperini} \& {Heggie}(1997)}]{vesp97a}
{Vesperini}, E., \& {Heggie}, D.~C. 1997, \mnras, 289, 898,
  \dodoi{10.1093/mnras/289.4.898}

\bibitem[{{Waldman} {et~al.}(2011){Waldman}, {Sauer}, {Livne}, {Perets},
  {Glasner}, {Mazzali}, {Truran}, \& {Gal-Yam}}]{wald11}
{Waldman}, R., {Sauer}, D., {Livne}, E., {et~al.} 2011, \apj, 738, 21,
  \dodoi{10.1088/0004-637X/738/1/21}

\bibitem[{{Wang} {et~al.}(2019){Wang}, {Leigh}, {Sesana}, \& {Perna}}]{wang19a}
{Wang}, Y.-H., {Leigh}, N., {Sesana}, A., \& {Perna}, R. 2019, \mnras, 482,
  3206, \dodoi{10.1093/mnras/sty2866}

\bibitem[{{Wechsler} \& {Tinker}(2018)}]{wech18a}
{Wechsler}, R.~H., \& {Tinker}, J.~L. 2018, \araa, 56, 435,
  \dodoi{10.1146/annurev-astro-081817-051756}

\bibitem[{{Woosley} \& {Kasen}(2011)}]{wk11}
{Woosley}, S.~E., \& {Kasen}, D. 2011, \apj, 734, 38,
  \dodoi{10.1088/0004-637X/734/1/38}

\bibitem[{{Yuan} {et~al.}(2013){Yuan}, {Kobayashi}, {Schmidt}, {Podsiadlowski},
  {Sim}, \& {Scalzo}}]{yuan13a}
{Yuan}, F., {Kobayashi}, C., {Schmidt}, B.~P., {et~al.} 2013, \mnras, 432,
  1680, \dodoi{10.1093/mnras/stt591}

\bibitem[{{Yungelson}(2008)}]{yung08}
{Yungelson}, L.~R. 2008, Astronomy Letters, 34, 620,
  \dodoi{10.1134/S1063773708090053}

\bibitem[{{Zenati} {et~al.}(2019{\natexlab{a}}){Zenati}, {Bobrick}, \&
  {Perets}}]{zena19b}
{Zenati}, Y., {Bobrick}, A., \& {Perets}, H.~B. 2019{\natexlab{a}}, \mnras,
  submitted (arXiv:1908.10866)

\bibitem[{{Zenati} {et~al.}(2019{\natexlab{b}}){Zenati}, {Perets}, \&
  {Toonen}}]{zena19a}
{Zenati}, Y., {Perets}, H.~B., \& {Toonen}, S. 2019{\natexlab{b}}, \mnras, 486,
  1805, \dodoi{10.1093/mnras/stz316}

\end{thebibliography}
\end{document}